\documentclass[preprintnumbers,pre,amsmath,amssymb,superscriptaddress,twocolumn]{revtex4-2}

\usepackage{graphicx}% Include figure files
\usepackage{dcolumn}% Align table columns on decimal point
\usepackage{bm}% bold math

\usepackage[normalem]{ulem}
\usepackage{nicefrac}
\usepackage[dvipsnames]{xcolor}
\usepackage{multirow}
\usepackage{amsfonts}
\usepackage{mathtools} %coloneqq
\usepackage{bbold}
\usepackage[bookmarks=true, colorlinks=true, linkcolor=RoyalBlue, urlcolor=blue, citecolor=OliveGreen, pdftex, bookmarks=true, hyperindex=true]{hyperref}
\usepackage{capt-of}
\usepackage{float}
\usepackage{lipsum}
\usepackage{soul}

\usepackage{color}
\usepackage{tikz}

\newcommand{\cs}{c_{\mbox{\scriptsize s}}}
\newcommand{\taulb}{\tau_{\mbox{\tiny LB}}}
\newcommand{\area}{\chi}
\renewcommand{\vec}[1]{\boldsymbol{\bm{#1}}}

\begin{document}

% \preprint{APS/123-QED}

\title{Immersed boundary - lattice Boltzmann method for wetting problems}

%Elisa Bellantoni, Fabio Guglietta, Francesca Pelusi, Mathieu Desbrun, Kiwon Um, Mihalis Nicolaou, Nikos Savva, Mauro Sbragaglia
\author{Elisa Bellantoni}
\email{e.bellantoni@cyi.ac.cy}
\affiliation{%
Computation-based Science and Technology Research Center, The Cyprus Institute,
20 Konstantinou Kavafi Street, 
2121 Nicosia, Cyprus
}
\affiliation{%
Department of Physics \& INFN, Tor Vergata University of Rome, 
Via della Ricerca Scientifica 1, 
00133 Rome, Italy
}%
\affiliation{%
LTCI, T\'{e}l\'{e}com Paris, IP Paris,
19 Place Marguerite Perey, 
91120 Palaiseau, France
}%

\author{Fabio Guglietta}%
\affiliation{%
Department of Physics \& INFN, Tor Vergata University of Rome, 
Via della Ricerca Scientifica 1, 
00133 Rome, Italy
}%

\author{Francesca Pelusi}%
\affiliation{%
Istituto per le Applicazioni del Calcolo, CNR
- Via Pietro Castellino 111, 
80131 Naples, Italy
}%

\author{Mathieu Desbrun}
\affiliation{%
Inria Saclay and LIX, IP Paris, 
1 rue Honor\'{e} d'Estienne d'Orves, 
91120 Palaiseau, France
}%

\author{Kiwon Um}
\affiliation{%
LTCI, T\'{e}l\'{e}com Paris, IP Paris, 
19 Place Marguerite Perey, 
91120 Palaiseau, France
}%

\author{Mihalis Nicolaou}
\affiliation{%
Computation-based Science and Technology Research Center, The Cyprus Institute, 
20 Konstantinou Kavafi Street, 
2121 Nicosia, Cyprus
}

\author{Nikos Savva}%
\affiliation{%
Computation-based Science and Technology Research Center, The Cyprus Institute, 
20 Konstantinou Kavafi Street, 
2121 Nicosia, Cyprus
}
\affiliation{%
Department of Mathematics and Statistics, University of Cyprus, 
1 Panepistimiou Avenue,
2109 Nicosia, Cyprus
}

\author{Mauro Sbragaglia}%
\affiliation{%
Department of Physics \& INFN, Tor Vergata University of Rome, 
Via della Ricerca Scientifica 1, 
00133 Rome, Italy
}%

\date{\today}% It is always \today, today,
             
\begin{abstract}
We develop a mesoscale computational model to describe the interaction of a droplet with a solid. The model is based on the hybrid combination of the immersed boundary and the lattice Boltzmann computational schemes: the former is used to model the non-ideal sharp interface of the droplet coupled with the inner and outer fluids, simulated with the lattice Boltzmann scheme. We further introduce an interaction force to model the wetting interactions of the droplet with the solid: this interaction force is designed with the key computational advantage of providing a regularization of the interface profile close to the contact line, avoiding abrupt curvature changes that could otherwise cause numerical instabilities. The proposed model substantially improves earlier immersed boundary - lattice Boltzmann models for wetting in that it allows a description of an ample variety of wetting interactions, ranging from hydrophobic to hydrophilic cases, without the need for any pre-calibration study on model parameters to be used. Model validations against theoretical results for droplet shape at equilibrium and scaling laws for droplet spreading dynamics are addressed. 
\end{abstract}
%Use showkeys class option if keyword
\maketitle

%%%%%%%%%%%%%%%%%%%%%%%%%%%%%%%%%%%%%%%
\section{\label{sec:intro}Introduction}
%%%%%%%%%%%%%%%%%%%%%%%%%%%%%%%%%%%%%%%%
Understanding the interaction between a liquid droplet and a solid substrate is a 
complex multiscale problem living at the crossroads between physics, chemistry, and engineering. At macroscopic scales, the ability of a droplet to wet a solid substrate is primarily quantified by the equilibrium contact angle $\theta_{\text{eq}}$ via the celebrated Young's law~\cite{DeGennes85,DeGennes2004capillarity},
\begin{equation*}
    \cos \theta_{\text{eq}} = \frac{\sigma_{\text{sg}}-\sigma_{\text{sl}}}{\sigma}\ ,
\end{equation*}
where $\sigma$ is the surface tension between the liquid (l) and the ambient gas (g), while $\sigma_{\text{sl}}$ and $\sigma_{\text{sg}}$ represent the surface tensions between the solid (s) and the liquid and gas, respectively. When a droplet is deposited on a solid substrate, a spreading process follows until an equilibrium shape with contact angle $\theta_{\text{eq}}$ is achieved. This spreading process has been the subject of intense scrutiny over the years, including experiments, theory, and numerical simulations~\cite{DeGennes85,DeGennes2004capillarity,bonn2009wetting,snoeijer2013,andreotti2020}. In this landscape, numerical simulations are optimal analysis tools, capable of revealing complex features of wetting dynamics~\cite{snoeijer2013,andreotti2020,de1999dynamic,kusumaatmaja2006drop,sbragaglia2007spontaneous,ding2007inertial,kusumaatmaja2008anisotropic,blow2009imbibition,savva2010two,moradi2010roughness,chinappi2010intrinsic,wheeler2010modeling,savva2011dynamics,carlson2011dissipation,carlson2012universality,gross2014spreading,frank2015lattice,LegendreMaglio13,legendre2015comparison,baroudi2020effect,guo2020modeling,chen2020dynamic,du2021initial,saiseau2022near,Pelusi2023,liu2024transition}.
Many recent studies~\cite{hosoi2004peeling,lister2013viscous,Doudrick14,carlson2018fluctuation,andreotti2020,Ni2021,poulain2022elastohydrodynamics,saeter2024coalescence} feature problems with complex interfaces, with added physical richness in comparison to models that account for surface tension forces alone; hence, the development of novel/improved numerical methods displaying flexibility and computational efficiency in modeling complex interface physics for wetting is highly desirable. 

The hybrid immersed boundary (IB) - lattice Boltzmann (LB) method is well-suited for capturing complex interface dynamics~\cite{Kruger17}. In this approach, the interface is sharp and is usually represented with a mesh on which the desired interface properties are introduced. In addition, the method is based on the LB approach, which is a highly efficient technique in computational fluid dynamics~\cite{Kruger17,Succi18}. Recent studies have used the IB-LB approach to model soft particles with complex interfaces, featuring viscoelasticity~\cite{Guglietta2020,rezghi2022tank,liSimilarDistinctRoles2021,guglietta2020lattice,guglietta2021loading,rezghi2022lateral,guglietta2023suspensions} and interfacial viscosity~\cite{li2020finite,guglietta24analytical,liFiniteDifferenceMethod2019,Pelusi2023}; however, to the best of our knowledge, applications of IB-LB to wetting problems have been much less studied in the literature. In Ref.~\cite{Pelusi2023}, Pelusi and co-workers proposed a computational model based on the IB-LB method to simulate the wetting dynamics of coated droplets. In their work, the interaction between the droplet and the solid substrate has been modelled via a Lennard-Jones interaction, which depends on two parameters that must be pre-calibrated to obtain the correct equilibrium contact angle. Moreover, as the authors stated, their implementation can capture only large contact angles. 
These issues are addressed in the present study via a systematic analytical control of the static solutions, thus avoiding pre-calibration efforts and substantially improving the model applicability in various physical scenarios.

To facilitate comparison with literature results, we focus on the case of a droplet with surface tension $\sigma$ at the interface. On the one hand, we aim to design and implement an interaction force with the wall that allows controlled modeling of the equilibrium contact angle $\theta_{\text{eq}}$, i.e., without the need for any pre-calibration step; on the other hand, we aim to extend the applicability of the model to address a broad spectrum of contact angles, ranging from hydrophobic to hydrophilic cases. To validate our IB-LB method, we perform comprehensive comparisons against analytical results for droplet shapes at equilibrium and for droplet dynamics during spreading.\\ 

The paper is organized as follows: in Sec.~\ref{sec:NUMERICS}, we review the basic features of the IB-LB method and describe the way wetting interactions with a solid substrate are introduced in the model; in Sec.~\ref{sec:RESULTS}, we report on the results of numerical simulations for both droplet statics and dynamics, with comparisons against known analytical results; conclusions follow in Sec.~\ref{sec:CONCLUSIONS}.

%%%%%%%%%%%%%%%%%%%%%%%%%%%%%%%%%%%%%%%%%%%%%%%%%%%%%%%%
\begin{figure*}[th!]
    \centering
    \includegraphics[width=1.\textwidth]{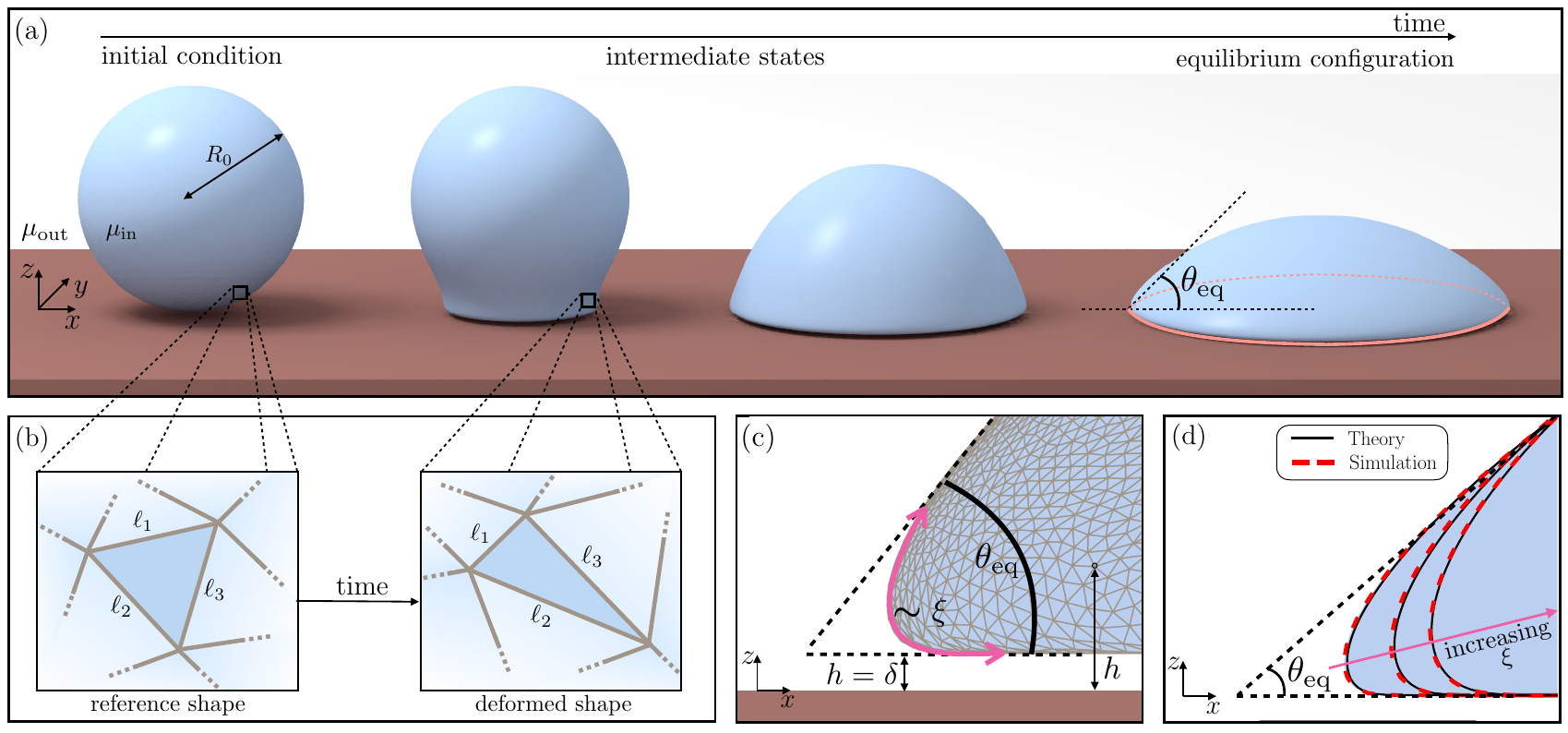}
    \caption{A sketch describing simulations of a droplet spreading on a flat substrate with the immersed boundary (IB) - lattice Boltzmann (LB) numerical method. Panel (a): macroscopic view of the spreading process allowing an initially spherical droplet with radius $R_0$ to attain its equilibrium configuration with contact angle $\theta_{\text eq}$. The droplet interface features a 3D triangular mesh that evolves in time [panel (b)] due to the action of surface tension and wetting forces [cf.~Eq.~\eqref{eq:force_split}]. The interface is further coupled (via the IB technique) with the dynamics of Newtonian viscous fluids (obtained with the LB technique) outside and inside the droplet, featuring dynamic viscosities $\mu_{\rm out}$ and $\mu_{\rm in}=\lambda \mu_{\rm out}$ respectively [cf.~Eq.~\eqref{eq:def_lambda}]. Panel (c): mesoscale regularization of the interface profile close to the contact line set by the wall-interaction term $\Pi(h)$, with $h$ the vertical distance of the interface from the wall [cf.~Eq.~\eqref{eq:disjoin-p}]. The change of the interface curvature from the outer region (spherical cap with contact angle $\theta_{\text eq}$) to the inner region (thin film with thickness $\delta$) is regulated by the lengthscale $\xi$, with increasing $\xi$ accommodating more gentle curvature changes [panel (d), see text for more details]. The static profiles are also controlled via analytical solutions [continuous lines in panel (d)].
    }
    \label{fig:concept_pic}
\end{figure*}
%%%%%%%%%%%%%%%%%%%%%%%%%%%%%%%%%%%%%%%%%%%%

%%%%%%%%%%%%%%%%%%%%%%%%%%
\section{Numerical method}\label{sec:NUMERICS}
%%%%%%%%%%%%%%%%%%%%%%%%%%
In this section, we provide details on the numerical method employed. The lattice Boltzmann (LB) method is used to resolve the fluids inside and outside the droplet, while the immersed boundary (IB) method is used to couple the droplet interface with the fluid. A wall-interaction force is further supplied to the IB scheme to model wettability. In Sec.~\ref{subsec:IBLB}, we recall the essential features of the IB-LB method; in Sec.~\ref{subsec:surf_tens}, we detail the modeling of surface tension forces; finally, in Sec.~\ref{subsec:wetting_interaction} we give specifics on the modeling of wettability.
%%%%%%%%%%%%%%%%%%%%%%%%%%%%%%%%%%%%%%%%%%%%%%%%%%%%%%%%%%%%%%%%%%%%%
\subsection{Immersed-boundary lattice Boltzmann (IB-LB) method}\label{subsec:IBLB}
%%%%%%%%%%%%%%%%%%%%%%%%%%%%%%%%%%%%%%%%%%%%%%%%%%%%%%%%%%%%%%%%%%%%%%%%%%
The IB-LB is a well-established method in the literature which is particularly suitable to reproduce flows and their coupling with deformable and complex interfaces~\cite{Kruger17,Guglietta2020,rezghi2022tank,liSimilarDistinctRoles2021,guglietta2020lattice,guglietta2021loading,guglietta2023suspensions,rezghi2022lateral,li2020finite,guglietta24analytical,liFiniteDifferenceMethod2019,taglientiReducedModelDroplet2023,Taglienti2024}. Fluid flows are described in terms of continuous fields depending on space (${\bm x}$) and time ($t$) via the continuity and Navier-Stokes equations:
\begin{gather}
\dfrac{\partial \rho}{\partial t} + \vec{\nabla}\cdot (\rho \vec{U})=0 \ , \label{eq:continuity} \\
\rho \left[ \dfrac{\partial \vec{U}}{\partial t} + (\vec{U} \cdot \vec{\nabla}) \vec{U} \right] = - \vec{\nabla}{p} + \mu \nabla^2{\vec{U}} + \vec{F} \ , \label{eq:NSE}
\end{gather}
with $\rho\!=\!\rho(\vec{x},t)$ being the fluid density, $\vec{U}\!=\!\vec{U}(\vec{x},t)$ the fluid velocity, $p\!=\!p(\vec{x},t)$  the fluid pressure, $\mu$ the dynamic viscosity, and $\vec{F}\!=\!\vec{F}(\vec{x},t)$ the force density. The LB method recovers this description by following a kinetic approach, which evolves in time a {\it discrete} probability distribution function (PDF), $f_i(\vec{x},t)$, corresponding to the probability density function of finding a fluid particle in the position $\vec{x}$ at time $t$ with lattice velocity $\vec{c}_i$. Space positions are discretized on a regular Cartesian lattice with a constant lattice spacing $\Delta x$ in all directions; correspondingly, time is discretized with a constant time step amplitude $\Delta t$. In LB models, a finite set of $Q$ velocity vectors ($i=0,\dots,Q-1$) is considered, along which the PDFs can stream~\cite{benzi1992lattice,Kruger17,Succi18}. Specifically, this work employs a D3Q19 velocity LB scheme featuring 19 velocity directions ($\vec{c}_i$) in a three-dimensional lattice. Velocity vectors are associated with statistical weights $\omega_i$. Details about the scheme are reported in Tab.~\ref{tab:D3Q19_set}. 
%%%%%%%%%%%%%%%%%%%%%%%%%%%%%%%%%%%%%%%%%%%%%%%%%%%%%%%%%%%%%%%%%%%%%%%
\bgroup\renewcommand{\arraystretch}{1.3}
\begin{table}[t!]
    \centering
    \begin{tabular}{lcc}
    \hline \hline
    $i$ & $\vec{c}_i$ & $\omega_i$ 
    \\\hline
    $0$   & $(0, 0, 0)$ & $ 1/3$ \\ [2pt]
    $1 - 6 $ & $(\pm \frac{\Delta x}{\Delta t}, 0, 0)$, $(0, \pm \frac{\Delta x}{\Delta t}, 0)$, $(0, 0, \pm \frac{\Delta x}{\Delta t})$ & $1/18$ \\ [2pt]
    $7 - 18$  & 
    $(\pm \frac{\Delta x}{\Delta t}, \pm \frac{\Delta x}{\Delta t}, 0)$, $(0, \pm \frac{\Delta x}{\Delta t}, \pm \frac{\Delta x}{\Delta t})$, $(\pm \frac{\Delta x}{\Delta t}, 0, \pm \frac{\Delta x}{\Delta t})$ 
    & $1/36$ \\ [2pt]
    \hline \hline
    \end{tabular}
    \caption{Velocity vectors $\vec{c}_i$ and weights $\omega_i$ for the D3Q19 LB scheme.}
    \label{tab:D3Q19_set}
\end{table}
\egroup
%%%%%%%%%%%%%%%%%%%%%%%%%%%%%%%%%%%%%%%%%%%%%%%%%%%%%%%%%%%%%%%%%%
The LB equations give the dynamic evolution of all the $f_i$, providing a version of the Boltzmann equation discretized in time, coordinate, and velocity space~\cite{benzi1992lattice,Kruger17,Succi18}:
\begin{equation}\label{eq:LBE}
% \small
    f_i(\vec{x} \!+\! \vec{c}_i \Delta t, t \!+\! \Delta t) - f_i(\vec{x}, t) \!=\! \Delta t \left[ \Omega_i(\vec{x},t) \!+\! S_i(\vec{x},t) \right]\, .
\end{equation}
Eq.~\eqref{eq:LBE} represents the sum of a streaming step (left-hand side), in which the populations are advected from one lattice node to another along the velocity vectors $\vec{c}_i$, and a collision process embedded in the collision term $\Omega_i(\vec{x},t)$. Streaming and collision steps are further supplemented by external body forces through a source term $S_i(\vec{x},t)$.  In this work, we employ the widely popular Bhatnagar-Gross-Krook (BGK) collision operator which is given by~\cite{Qian1992}
\begin{equation}
\Omega_i(\vec{x},t) = - \frac{1}{\taulb}\left[ f_i(\vec{x}, t) - f_{i}^{(\text{eq})}(\vec{x}, t) \right]\, ,
\end{equation}
where $\taulb$ is the characteristic relaxation time and $f_{i}^{(\text{eq})}(\vec{x}, t)$ is the Maxwell-Boltzmann equilibrium distribution function that depends on $(\vec{x},t)$ via the density and velocity field, i.e., $f_{i}^{(\text{eq})}(\vec{x}, t)\!=\!f_{i}^{(\text{eq})}(\rho(\vec{x}, t),\vec{U}(\vec{x}, t))$. The equilibrium distribution is taken as the Taylor expansion of the Maxwell-Boltzmann distribution
\begin{equation}
f_{i}^{(\text{eq})}(\rho, \vec{U}) = \omega_i \rho \left[ 1 + \dfrac{ \vec{U} \cdot \vec{c}_i}{ \cs^2 } + \dfrac{ (\vec{U} \cdot \vec{c}_i)^2}{ 2\cs^4 } - \dfrac{ \vec{U} \cdot \vec{U}}{2 \cs^2 } \right]\, .
\end{equation}
Here, $\cs$ is the lattice speed of sound, which corresponds to $\cs\!=\!\Delta x / (\sqrt{3}\Delta t)$ in the case of the D3Q19 velocity scheme. Higher-order collision operators are possible, but they elude the scope of the present work. The contribution of the force density $\vec{F}(\vec{x},t)$ is encoded in the source term $S_i(\vec{x},t)$, which, in turn, depends on $(\vec{x},t)$ via the velocity and the force density, i.e., $S(\vec{x},t)=S(\vec{U}(\vec{x},t),\vec{F}(\vec{x},t))$, according to the Guo's forcing scheme~\cite{Guo02}
\begin{equation}
    S_i(\vec{U},\vec{F}) = \left( 1 - \frac{\Delta t}{2 \taulb} \right) \omega_i \left[ \dfrac{\vec{c}_i - \vec{U}}{ \cs^2 } + \left(\dfrac{ \vec{U} \cdot \vec{c}_i}{ \cs^4 }\right) \vec{c}_i \right] \cdot \vec{F}\, .
\end{equation}
The hydrodynamic variables are recovered by taking the moments of the PDF~\cite{Guo02,Kruger17}:
\begin{gather}
\rho(\vec{x},t) = \sum_i f_i(\vec{x},t) \label{eq:hydro_var1} \, , \\ 
\vec{U}(\vec{x},t) = \frac{1}{\rho(\vec{x},t)}\sum_i \vec{c}_i f_i(\vec{x},t) + \dfrac{\vec{F}(\vec{x},t) \Delta t}{2\rho(\vec{x},t)} \, ,\label{eq:hydro_var2}
\end{gather}
where the equation for the fluid velocity $\vec{U}$ contains the half-force correction~\cite{Kruger17,Guo02}. The continuity and the Navier-Stokes equations (cf.~Eqs.~\eqref{eq:continuity}-\eqref{eq:NSE}) are recovered for the hydrodynamic variables in Eqs.~\eqref{eq:hydro_var1}-\eqref{eq:hydro_var2}, with dynamic viscosity set by $\mu \!=\!\rho \cs^2 (\taulb - \Delta t /2)$, while the pressure is given by $p\!=\!\cs^2\rho$ \cite{Kruger16}.
To account for the varying fluid viscosity inside and outside the droplet, we consider the LB relaxation time as a function of both space and time: $\taulb=\taulb(\vec{x},t)$. At each time step, using a ray tracing algorithm already employed in previous works \cite{taglientiReducedModelDroplet2023,Taglienti2024,guglietta24analytical}, we identify the lattice nodes located inside and outside the closed interface and set them to the values $\taulb=\taulb^{\text{in}}$ and $\taulb^{\text{out}}$, respectively.
Correspondingly, we define the viscosity ratio $\lambda$ as the ratio between the inner dynamic viscosity $\mu_{\text{in}}\!=\!\rho \cs^2 (\taulb^{\text{in}} - \Delta t /2)$ and the outer one $\mu_{\text{out}}\!=\!\rho \cs^2 (\taulb^{\text{out}} - \Delta t /2)$:
\begin{equation}\label{eq:def_lambda}
\lambda=\frac{\mu_{\text{in}}}{\mu_{\text{out}}}.
\end{equation}
To model the droplet interface, we introduce a set of Lagrangian nodes arranged in a 3D triangular mesh (see Fig.~\ref{fig:concept_pic}). The positions of the nodes, $\vec{q}_{j}(t)$ ($j=1,\dots, N$, where $N$ is the number of nodes), change over time following the droplet motion, deforming the $N_t$ triangles of the mesh. The interaction between the fluid and the droplet interface relies on a two-way coupling between the Eulerian lattice and the Lagrangian nodes: at every time step, nodal forces $\vec{\varphi}_j$ are evaluated on every mesh node and spread onto the fluid nodes to obtain the local force density ${\bm F}$; at the same time, a no-slip condition is enforced by requiring the velocity of the Lagrangian nodes, $\dot{\vec{q}}_j(t)$, to match the surrounding fluid velocity $\vec{U}$. The fluid-structure coupling is then obtained as:
\begin{gather}\label{eq:force_spread}
    \vec{F}(\vec{x},t) = \sum_{j} \vec{\varphi}_j(t) \Delta(\vec{q}_j(t)-\vec{x}) \ ,  \\ \label{eq:vel_interpolation}
    \dot{\vec{q}}_j(t) = \sum_{\vec{x}} \vec{U}(\vec{x},t) \Delta(\vec{q}_j(t) - \vec{x})\Delta x^3\, . 
\end{gather}
The above is implemented by using a four-point interpolation stencil, encoding a discrete approximation of the Dirac delta function $\delta(\vec{x})$ as $\Delta(\vec{x})\!\coloneqq\!\Psi(x)\Psi(y)\Psi(z)/\Delta x^3$, where $\Psi(x)$ is given by~\cite{Kruger17}
\small{
\begin{equation}
\Psi(x) = 
    \begin{cases}\displaystyle
        \frac{3}{8}-\frac{1}{4}\frac{\vert x\vert}{\Delta x} +\frac{1}{4} \sqrt{\frac{\vert x\vert}{\Delta x} -\frac{x^2}{\Delta x^2} +\frac{1}{4} },  & \dfrac{\vert x\vert}{\Delta x}\le 1 \\[1.1em]
        \displaystyle\frac{5}{8}-\frac{1}{4}\frac{\vert x\vert}{\Delta x} - \frac{1}{4}\sqrt{3\frac{\vert x\vert}{\Delta x}-\frac{x^2}{\Delta x^2}-\frac{7}{4}}, & 1 \le \dfrac{\vert x\vert}{\Delta x}\le 2 \\[1.1em] 
        0, & \dfrac{\vert x\vert}{\Delta x}\ge2
    \end{cases}
\end{equation}
}
and, analogously, $\Psi(y)$ and $\Psi(z)$.
Finally, at every time step, the positions of the Lagrangian nodes are updated using a single-step forward Euler scheme:
\begin{equation}
    \vec{q}_j(t + \Delta t) = \vec{q}_j(t) + \dot{\vec{q}}_j(t) \Delta t\, .
\end{equation}
Surface tension and wettability forces are encoded in the force density $\vec{F}$ (see Eq.~\eqref{eq:NSE}) which is localized at the interface. Thus, we split its contributions into two parts: a surface tension force ($\vec{F}_{\sigma}$) and a wetting force ($\vec{F}_{\Pi}$). In formulas, this means that:
\begin{equation}\label{eq:force_split}
\begin{split}
\vec{F}({\bm x},t) & =\left[\vec{F}_{\sigma}+\vec{F}_{\Pi} \right]\delta({\bm x}-{\bm r}) = \\
& = -\left[\sigma \bm{\hat{n}} \, ({\bm \nabla} \cdot \bm{\hat{n}}) + \Pi(h) \bm{\hat{n}} \right]\delta({\bm x}-{\bm r}) 
\end{split}
\end{equation}
where $\bm{\hat{n}}$ is the normal at the interface pointing out of the droplet, and $\Pi(h)$ represents an interaction term that depends on the distance $h$ of the interface node from the wall (see Fig.~\ref{fig:concept_pic}(c)). The latter represents an important computational ingredient of this work, since it controls the droplet wettability. Its implementation,  whose details are provided in Sec.~\ref{subsec:wetting_interaction}, marks a key distinction from the approach in Ref.~\cite{Pelusi2023}, where the wetting dynamics is driven by a Lennard–Jones force orthogonally-oriented with respect to the wall. By adopting Eq.~\eqref{eq:force_split} and employing a proper definition of the interaction term $\Pi(h)$ (see Sec.~\ref{subsec:wetting_interaction}), we overcome the limitations reported in Ref.~\cite{Pelusi2023} regarding the impossibility to achieve small contact angles. As discussed below, Eq.~\eqref{eq:force_split} provides greater analytical control over the solution.
In the following sections, we will report details on how the surface tension is implemented in our model and on the choice of the interaction term $\Pi(h)$.
%%%%%%%%%%%%%%%%%%%%%%%%%%%%%%%%%%%%%%%%%%%%%%%%%%%%%%%%%%%%%%%%%%%%%
\subsection{Modelling surface tension forces}\label{subsec:surf_tens}
%%%%%%%%%%%%%%%%%%%%%%%%%%%%%%%%%%%%%%%%%%%%%%%%%%%%%%%%%%%%%%%%%%%%%
To simulate a droplet with a given surface tension $\sigma$, we follow the approach presented in Ref.~\cite{Pelusi2023} and 
rely on consolidated results from elasticity theory~\cite{green1960large,barthes1981time,dimitrienko2010nonlinear}, here briefly reviewed. Let ${\bm X}(t)$ be the position of a surface element, and ${\bm X_0}$ the position of the same element in some reference configuration (e.g., in the configuration at rest). The deformation gradient tensor is given by: 
\begin{equation}
\bm C(t)=\frac{\partial {\bm X}(t)}{\partial{\bm X_0}}\ .
\end{equation}
Since we are interested in interfacial modeling, we consider the following projection onto the surface of the droplet: 
\begin{equation}\label{eq:surface_deformation}
\bm D(t) = {\bm P}(t)\cdot{\bm C}(t)\cdot{\bm P(t=0)} \ ,
\end{equation}
where $\bm P(t)=\left[\mathbb{1}-\bm{\hat{n}}(t)\otimes \bm{\hat{n}}(t)\right]$ is a projection operator, with $\mathbb{1}$ being the 3D identity tensor~\cite{barthes1981time}. One can define two strain invariants starting from $\bm D$, i.e.,
\begin{subequations}\label{eq:strain_inv}
\begin{equation}
    a =
    \frac{1}{2}\log\left\{
    \frac{1}{2}\!\left[\mbox{tr}(\bm D\cdot\bm D^T)\right]^2\!\!-\,
    \frac{1}{2}\mbox{tr}\!\left[(\bm D\cdot\bm D^T)^2\right]\right\} \ ,\vspace*{-3mm}
\end{equation}
\begin{equation}
    b =
    \frac{1}{2}\mbox{tr}(\vec{D}\cdot\vec{D}^T)-1 \ ,
    \end{equation}
\end{subequations}
that can be used to write the interface stress $\bm \Sigma$ and the force $\bm F_{\sigma}$ exerted by the interface~\cite{barthes1981time}:
\begin{equation}
    \bm \Sigma = e^{-a}\left[\frac{\partial w}{\partial a}\bm P(t) +\frac{\partial w}{\partial b}\bm D\cdot\bm D^T \right] \ ,\vspace*{-3mm}
\end{equation}
\begin{multline}\label{eq:force_bb}
    \bm F_{\sigma} = \bm P \cdot \bm\nabla\cdot \bm\Sigma = -e^{-a}\left(\frac{\partial w}{\partial a}+\frac{\partial w}{\partial b}\right)\bm{\hat{n}}\bm\nabla\cdot\bm{\hat{n}}+ \\
    + \bm P\cdot\bm\nabla\left[e^{-a}\left(\frac{\partial w}{\partial a}+\frac{\partial w}{\partial b}\right)\right]+\\
    +\bm P \cdot\bm\nabla\cdot\left[e^{-a}\frac{\partial w}{\partial b}(\bm D\cdot\bm D^T-\mathbb{1})\cdot(\mathbb{1} -\bm{\hat{n}} \otimes \bm{\hat{n}})\right]\ ,
\end{multline}
with $w$ being the strain energy function that characterizes the mechanical response of the interface. We can recognise the surface tension coefficient $\sigma$ in the first term of Eq.~\eqref{eq:force_bb} as
\begin{equation}
\sigma \equiv e^{-a}\left(\frac{\partial w}{\partial a}+\frac{\partial w}{\partial b}\right)\ .
\end{equation}
As shown in Ref.~\cite{barthes1981time}, one can choose the strain energy for a droplet with surface tension $\sigma$, such that $e^{-a}\frac{\partial w}{\partial a} = \sigma$ and $\frac{\partial w}{\partial b}=0$, thus obtaining:
\begin{equation}\label{eq:energy_drop}
    w(a) = \sigma e^{a}\ .
\end{equation}
Eq.~\eqref{eq:force_bb} therefore becomes:
\begin{equation}
\bm F_{\sigma} = -\sigma\bm{\hat{n}} (\bm \nabla\cdot\bm{\hat{n}}).
\end{equation}
Note that the second term of Eq.~\eqref{eq:force_bb} is zero because $\sigma$ is constant. We remark that the force in Eq.~\eqref{eq:force_bb} is defined and computed at the droplet interface before being spread to the fluid via the IB-LB coupling described in the previous section, hence resulting in the first term of the right-hand side of Eq.~\eqref{eq:force_split}. The numerical implementation of the surface tension force follows finite-element-method-like technique whose details are given in Ref.~\cite{KrugerPhDthesis12}, here briefly sketched. On each triangle of the triangulated mesh, whose vertices' positions are $\bm q_{1,2,3}(t)$, we define the displacement vectors, $\bm V_{1,2,3}(t) = \bm q_{1,2,3}(t) - \bm q_{1,2,3}(t=0)$, which are used to compute the displacement: 
\begin{equation}\label{eq:V}
    \bm V(t) = N_1\bm V_1 + N_2\bm V_2 + N_3\bm V_3 \ ,
\end{equation}
with $N_{1,2,3}(x,y) = a_{1,2,3} x+b_{1,2,3}y+c_{1,2,3}$ being the shape functions and whose coefficients are reported in Ref.~\cite{KrugerPhDthesis12}. 
Eq.~\eqref{eq:V} can be used to formulate the discrete version of Eq.~\eqref{eq:surface_deformation}:
\begin{equation}
    \bm{\mathcal{D}}(t) = {\bm I} + \vec{\nabla}\bm V(t) \ ,
\end{equation}
where $\vec{\nabla}\bm V$ is a 2D second-rank tensor, and  ${\bm I}$ is the 2D identity tensor.
We then use $\bm{\mathcal{D}}$ to compute the strain invariants on each triangle (see Eqs.~\eqref{eq:strain_inv}), which are used to evaluate the strain energy $w$ [see Eq.~\eqref{eq:energy_drop}]. Finally, the contribution to the force density on each node of the mesh given by the surface tension at the interface is:
\begin{equation}
    {\bm \varphi}_{\sigma,j} = - \frac{\partial w}{\partial{\bm V_j}}\ .
\end{equation}
%%%%%%%%%%%%%%%%%%%%%%%%%%%%%%%%%%%%%%%%%
\subsection{Modelling wetting properties}\label{subsec:wetting_interaction}
%%%%%%%%%%%%%%%%%%%%%%%%%%%%%%%%%%%%%%%%%
Earlier studies~\cite{Gohl18} used the IB approach to investigate dynamic contact angles on surfaces with complex wettability. In that approach, the contact angle is explicitly imposed as a boundary condition, which enables the treatment of surfaces with complex features. However, it requires the use of interface normals, which is particularly challenging near boundaries. In contrast, in our LB implementation of wetting properties, we avoid challenges associated with accurately defining the interface normals near solid boundaries, and we regularize the interface curvature in such a way that the desired contact angle emerges as the large-scale limit of some inner mesoscale interaction model. This is achieved via the introduction of the interaction term $\Pi(h)$ in Eq.~\eqref{eq:force_split} as follows~\cite{Chamakos_2013,Karapetsas_2016,du2021initial}
\begin{equation}\label{eq:disjoin-p}
\Pi(h) = A\left[\left( \frac{\xi}{h}\right)^n-\left( \frac{\xi}{h}\right)^m\right] \ ,
\end{equation}
where $n$ and $m$ ($n\!>\!m$) are parameters regulating the effective range of the interaction, $A$ is constant, and $h$ measures the (vertical) distance from the solid surface [see Fig.~\ref{fig:concept_pic}(c)]. The {\it regularizing lengthscale} $\xi$ is tunable in the model and sets the characteristic lengthscale that regulates interface curvature changes when the outer interface profile approaches solid walls [see Fig.~\ref{fig:concept_pic}(c)]. Regarding the numerical implementation, the interaction term $\Pi(h)$ is computed on each Lagrangian node of the mesh representing the interface.
For each node $j$, we calculate the corresponding wetting nodal force, $\bm \varphi_{{\scriptscriptstyle\Pi},_j}$, which contributes to the IB method [see Eq.~\eqref{eq:force_spread}] as 
\begin{equation}
    \bm \varphi_{{\scriptscriptstyle\Pi},_j} = - \frac{\area_j}{3}\Pi(h) \bm{\hat{n}}_j\ , 
\end{equation}
where $\bm{\hat{n}}_j$ is taken as the average of the normal vectors of all faces that share the $j$-th node, while $\area_j$ represents the area of the mesh associated with the $j$-th node. The factor $1/3$ accounts for each triangle area being shared by three vertices. This renormalization using the nodal area is necessary to ensure robustness to mesh discretization. Similar to the surface tension forces, the wetting force is spread to the fluid via the IB method, resulting in the second term of the right-hand side of Eq.~\eqref{eq:force_split}.
Let us now discuss the choice of the parameters $m$, $n$, $A$, and $\xi$ of the wall-interaction force. From the balance between the surface tension forces with the wetting forces at the interface one obtains 
\begin{equation}\label{eq:laplace-pi}
\Delta p= \sigma\, {\bm \nabla} \cdot \bm{\hat{n}}+\Pi(h)\ ,
\end{equation}
where $\Delta p$ is the pressure jump across the droplet interface. In Appendix~\ref{appendix:setting_angle}, we show that the interface profile  displays a wedge-like structure near the wall, which becomes more prominent as the micro-scale $\xi$ becomes much smaller than the droplet radius, $R_0$, noting that the infinite, wedge-like assumption becomes less relevant for smaller contact angles (see Appendix~\ref{appendix:DROPLET_ASYMPTOTICS}). This wedge is characterized by an angle $\theta_{\text{eq}}$, which is related to $\sigma$, $A$, $n$, and $m$ as follows~\cite{Chamakos_2013,Karapetsas_2016,du2021initial}:
\begin{equation}\label{eq:A_choice}
A =\sigma \frac{(m-1)(n-1)}{(n-m)\xi}(1+\cos\theta_{\text{eq}}).
\end{equation}
Eq.~\eqref{eq:A_choice} allows a controlled use of the parameters $A$, $m$, $n$, and $\xi$ with a precise link with the equilibrium contact angle $\theta_{\text{eq}}$, without the need of any pre-calibration step on model parameters, as done in Ref.~\cite{Pelusi2023}. Moreover, as discussed in detail in the next section [see also Fig.~\ref{fig:concept_pic}(d)], for a given $\theta_{\text eq}$, the lengthscale $\xi$ can be further used to achieve the desired level of interface profile regularization close to the contact line.

The use of the wall-interaction force of the form given in Eq.~\eqref{eq:disjoin-p} echoes the disjoining-pressure term which is commonly invoked to account for the inter-molecular interactions of thin films near a substrate~\cite{Israelachvili2011intermolecular}, allowing the formation of a thin pre-wetting film ahead of the contact line in pseudo-partial wetting scenarios~\cite{Brochard1991spreading,DeGennes2004capillarity,Eggers2005contact}. Here, as in relevant works in the literature~\cite{Chamakos_2013,Karapetsas_2016,du2021initial}, the Lennard--Jones-like interaction term $\Pi(h)$ is used to model strong repulsions at short distances between the droplet and the substrate, and attraction at intermediate distances. 
In this manner, $\Pi(h)$ primarily impacts the contact angle and the shape of the interface near the substrate. Importantly, its presence regularizes the curvature of the droplet interface, suppressing any stresses that would have resulted in the vicinity of a moving contact line. Indeed, dealing with abrupt curvature changes would introduce strongly localized forces in the interface model, which, in turn, could cause numerical instabilities when the interface is coupled with the fluid via the IB method (see Sec.~\ref{subsec:IBLB})~\cite{Kruger17}.
We emphasize that our approach in modeling wetting is not primarily aimed at establishing a precise link between $\Pi(h)$ and the underlying molecular physics. Instead, we wish to utilize an efficient and straightforward computational strategy (inspired by physics) that allows smooth curvature changes over the lengthscale $\xi$ near the contact line. Yet, we remark that the macroscopic features of wetting (i.e., when $\xi/R_0 \ll 1$) are reproduced in our quantitative model validations, both for droplet statics (see Sec.~\ref{subsec:STATICS}) and spreading (see Sec.~\ref{subsec:DYNAMICS}). 

%%%%%%%%%%%%%%%%%%%%%%%%%%%%%%%%%%%%%%%%%%%%%%%%%%%%%%%%
\begin{figure*}[th!]
    \centering
    \includegraphics[width=0.85\textwidth]{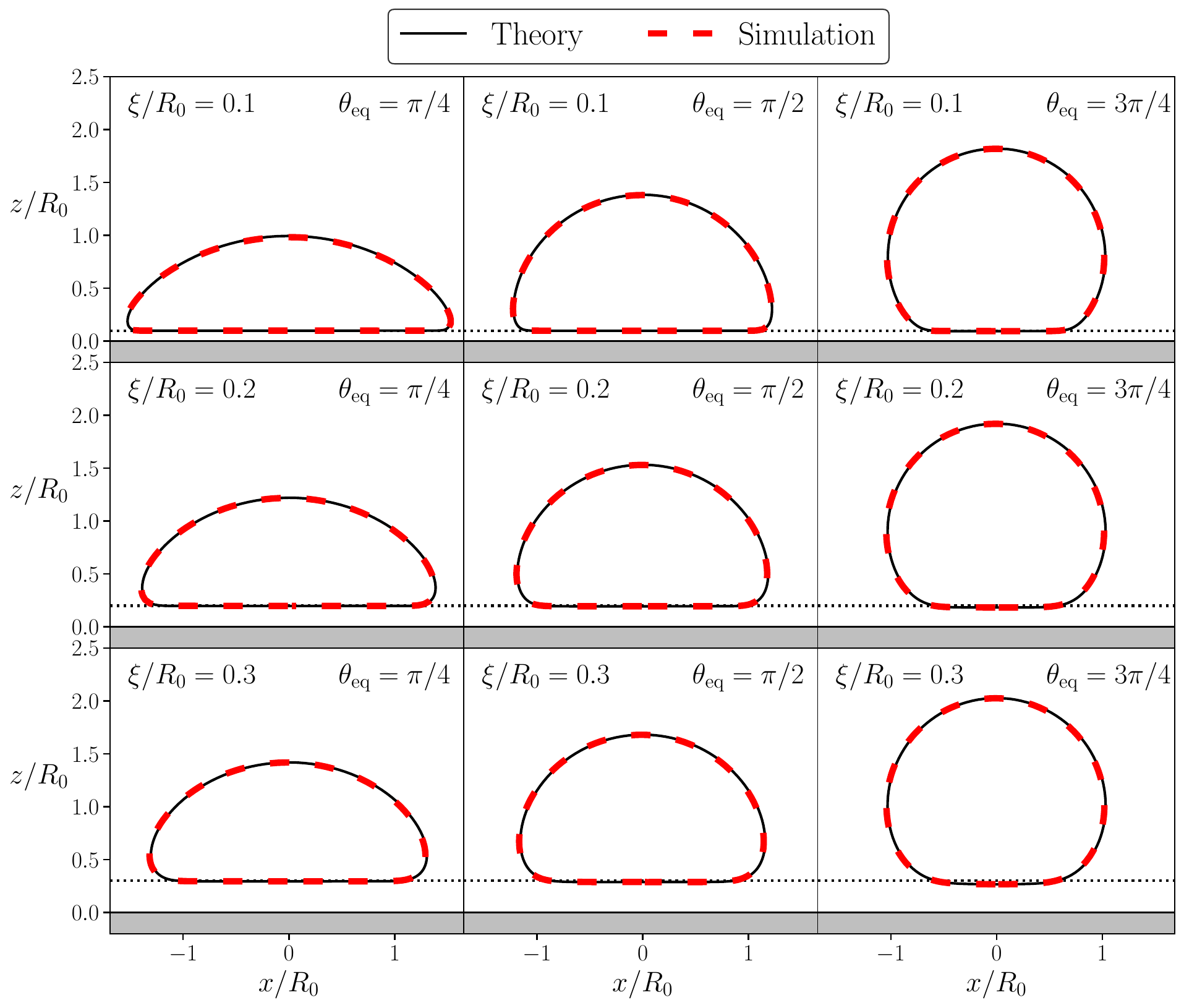}
    \caption{
    Comparison of droplet profiles at equilibrium between the equilibrium solution obtained by solving Eq.~\eqref{eq:laplace-pi} (solid black lines) and the simulated results (dashed red lines) for different combinations of equilibrium contact angle, $\theta_{\text{eq}}$ and normalized regularization lengthscale $\xi/R_0$ (different panels): $\theta_{\text{eq}}$ increases from left to right, while $\xi/R_0$ increases from top to bottom. The dotted line represents the height $z/R_0=\xi/R_0$ [see Eq.~\eqref{eq:disjoin-p}].}
    \label{fig:drops_statics}
\end{figure*}
%%%%%%%%%%%%%%%%%%%%%%%%%%%%%%%%%%%%%%%%%%%%

%%%%%%%%%%%%%%%%%%%%%%%%%%%
\section{Numerical results}\label{sec:RESULTS}
%%%%%%%%%%%%%%%%%%%%%%%%%%%
In this section, we discuss numerical simulation results to validate the implementation of the IB-LB model. First, we consider the droplet at equilibrium and compare the droplet's shape obtained from numerical simulations with the corresponding theoretical prediction (Sec.~\ref{subsec:STATICS}). Then, in Sec.~\ref{subsec:DYNAMICS}, we consider the spreading dynamics of the droplet towards equilibrium shapes.\\
The numerical simulations have been carried out in a 3D domain, with edges of length $\displaystyle L_x$, $L_y$, and $L_z$. In all simulations, we set $\Delta x=\Delta t = 1$, hence they will be dropped hereafter. The boundary conditions are taken to be periodic along the $x$ and $y$ directions [see Fig.~\ref{fig:concept_pic})] while the no-slip boundary condition is enforced at the top and bottom boundaries using a bounce-back approach~\cite{Kruger17}. 
Throughout this work, the initial density was always set to unity and the exponents in Eq.~\eqref{eq:disjoin-p} were taken to be $(n,m)\!=\!(6,3)$ as attested in related studies \cite{Zitz19}.
Simulations ran on NVIDIA Ampere A100 64GB graphics processing units (GPUs), with a simulation time for each spreading process lasting between 0.5 and 12 hours, depending on simulation parameters.
%%%%%%%%%%%%%%%%%%%%%%%%%%%%%%%%%%%%%%%%%%
\subsection{Statics}\label{subsec:STATICS}
%%%%%%%%%%%%%%%%%%%%%%%%%%%%%%%%%%%%%%%%%%%
We started our investigations by analyzing the droplet shapes in equilibrium conditions. Numerical simulations were performed in a box of dimension $\displaystyle L_x \!\times\! L_y \!\times\! L_z \!=\! 80 \!\times\! 80 \!\times\! 60$. To reach equilibrium, we start with an equilibrated spherical droplet, with initial radius $R_0\!=\!20$ modelled with $N_t\!=\!40\,000$ triangles, which is placed close to the substrate, at a distance where the interaction term $\Pi(h)$ is non-negligible. This way, the droplet starts to spread [see Fig.~\ref{fig:concept_pic}(a)] until the equilibrium state is achieved. The viscosity ratio between the inner and outer phases is $\lambda\!=
\!1$. We considered different equilibrium contact angles, $\theta_{\text{eq}}\!\in\!\{\pi/4$, $\pi/2$, $3 \pi/4\}$, and different values of the normalized regularization lengthscale, $\xi/R_0\!\in\!\{0.1,0.2,0.3\}$. 
\par
The equilibrium droplet shape is extracted from numerical simulations and compared with theoretical predictions. 
In particular, under the assumption of an axisymmetric droplet shape, one can derive a differential equation for the profile of the surface of the droplet, which is described in spherical coordinates in terms of the radial distance from the origin with a profile of the form $r=G(\phi)$,  with $\phi$ being the polar angle formed with the $z$-axis. 
For a schematic, derivation and implementation details of the differential equation for the droplet shape [Eq.~\eqref{eq:laplace-pi}], see Appendix~\ref{appendix:droplet_shape}; for further commentary on its asymptotic structure see Appendix~\ref{appendix:DROPLET_ASYMPTOTICS}.

In Fig.~\ref{fig:drops_statics}, we report a comparison between the theoretical droplet shape obtained by solving Eq.~\eqref{eq:laplace-pi} (solid black lines) and the simulated ones (dashed red lines) 
extracted by taking a slice over a plane orthogonal to the $y$-axis and passing through the droplet's center of mass. Overall, the agreement between the simulated droplet shapes and the theoretical predictions is excellent for all combinations of $\xi/R_0$ and $\theta_{\mathrm{eq}}$. 
To delve deeper into the matter, we computed the error between the theoretical radial profile $G_{\mbox{\tiny theo}}(\phi)$ and the simulated one $G_{\mbox{\tiny sim}}(\phi)$ by averaging on the azimuthal angle $\phi$:
\begin{gather}\label{eq:L2-error_equilibrium}
    {\mathrm{E}_{L_2}}= \sqrt{ \frac{ \int_0^{\pi} |G_{\mbox{\rm \tiny sim}}(\phi)-G_{\mbox{\rm \tiny theo}}(\phi)|^2 \mathrm{d}\phi}{\int_0^{\pi} |G_{\mbox{\rm \tiny theo}}(\phi)|^2 \mathrm{d}\phi} }\, .
\end{gather}
Here, the integration domain is restricted to $\phi \in [0,\pi]$ because of the axial symmetry of the problem. 
%%%%%%%%%%%%%%%%%%%%%%%%%%%%%%%%%%%%%%%%%%%%%%%%%%%%%%%%%%%%%
\begin{figure}[t!]
    \centering
    \includegraphics[width=0.5\textwidth]{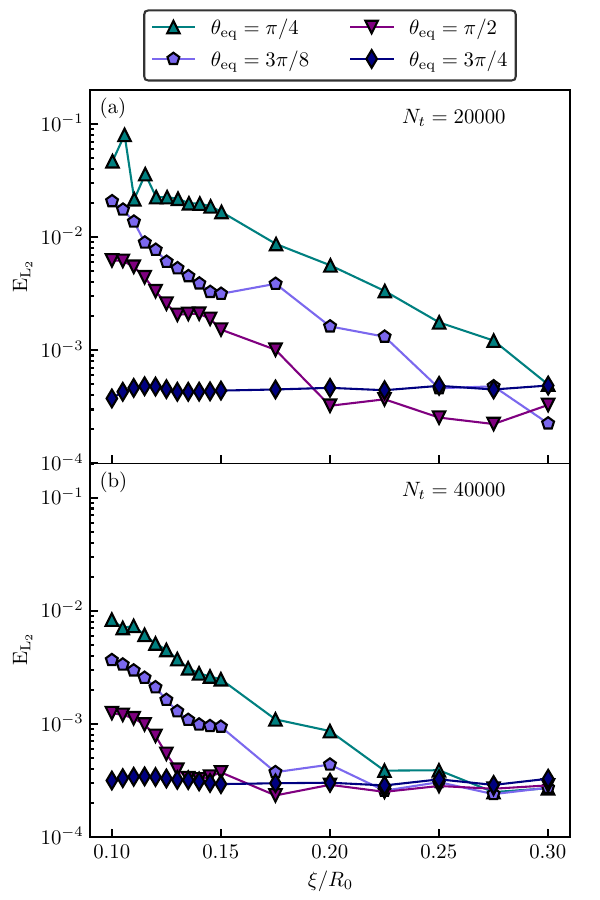}
    \caption{
    Error $\mathrm{E}_{L_2}$ between IB-LB simulations and theoretical prediction (cf.~Eq.~\eqref{eq:L2-error_equilibrium}) as a function of the normalized regularization lengthscale, $\xi/R_0$, for different values of $\theta_{\text eq}$ (different symbols/colors). We consider different realizations for the number of triangles on the droplet interface: $N_t\!=
    \!20\,000$ (top panel) and $N_t\!=\!40\,000$ (bottom panel).}
    \label{fig:drops_statics_error}
\end{figure}
%%%%%%%%%%%%%%%%%%%%%%%%%%%%%%%%%%%%%%%%%%%%%%%%%%%%%%%%%%%%%
Fig.~\ref{fig:drops_statics_error} shows the error $\mathrm{E}_{L_2}$ as a function of $\xi/R_0$ for different values of $\theta_{\text{eq}}$. 
There, $G{\mbox{\tiny sim}}(\phi)$ is extracted by considering the radial distance of every Lagrangian node from the droplet's centroid.
The radius $R_0$ is kept fixed to $R_0\!=\! 20$. To evaluate the influence of interface resolution, we performed numerical simulations with different numbers of triangles: $N_t \!=\! 20\,000$ [Fig.~\ref{fig:drops_statics_error}(a)] and $N_t \!=\! 40\,000$ [Fig.~\ref{fig:drops_statics_error}(b)]. In Fig.~\ref{fig:drops_statics_error}(a), the error $\mathrm{E}_{L_2}$ is generally small, decreasing to approximately $0.001$ as $\xi/R_0$ increases. For the largest value of the equilibrium contact angle ($\theta_{\text{eq}} \!=\! 3\pi/4$), the error is independent of $\xi/R_0$. Conversely, for smaller values of $\theta_{\text{eq}}$, the error increases as $\xi/R_0$ decreases, becoming pronounced around a critical value $(\xi/R_0)^*$. The critical value $(\xi/R_0)^*$ is larger for smaller values of $\theta_{\text{eq}}$: for $\theta_{\text{eq}} \!=\! \pi/4$, the error reaches a peak value of approximately $\mathrm{E}^{\text{peak}}_{L_2} \simeq 0.1$. These effects are likely tied to the discretization of the droplet interface. By comparing Fig.~\ref{fig:drops_statics_error}(a) and Fig.~\ref{fig:drops_statics_error}(b), it is clear that higher resolution leads to a transition to smaller errors occurring at lower values of $(\xi/R_0)^*$. Additionally, the peak error, $\mathrm{E}^{\text{peak}}_{L_2}$, is reduced for smaller values of $\theta_{\text{eq}}$ when a finer discretization is used. Such a resolution study reveals that $N_t \!=\! 40\,000$ provides a good resolution, as the maximum error $\mathrm{E}^{\text{peak}}_{L_2} \simeq 0.01$ for the smallest $\theta_{\text{eq}}$ considered. In contrast, $N_t \!=\! 20\,000$ does not achieve good convergence, as evidenced by higher errors and oscillations for small values of $\xi/R_0$. We therefore performed all simulations for the statics with $N_t\!=\!40\,000$.
%%%%%%%%%%%%%%%%%%%%%%%%%%%%%%%%%%%%%%%%%%%%%
\subsection{Dynamics}\label{subsec:DYNAMICS}
%%%%%%%%%%%%%%%%%%%%%%%%%%%%%%%%%%%%%%%%%%%%
\begin{figure*}[th!]
    % \centering
    \makebox[\textwidth][c]{\includegraphics[width=1.2\textwidth]{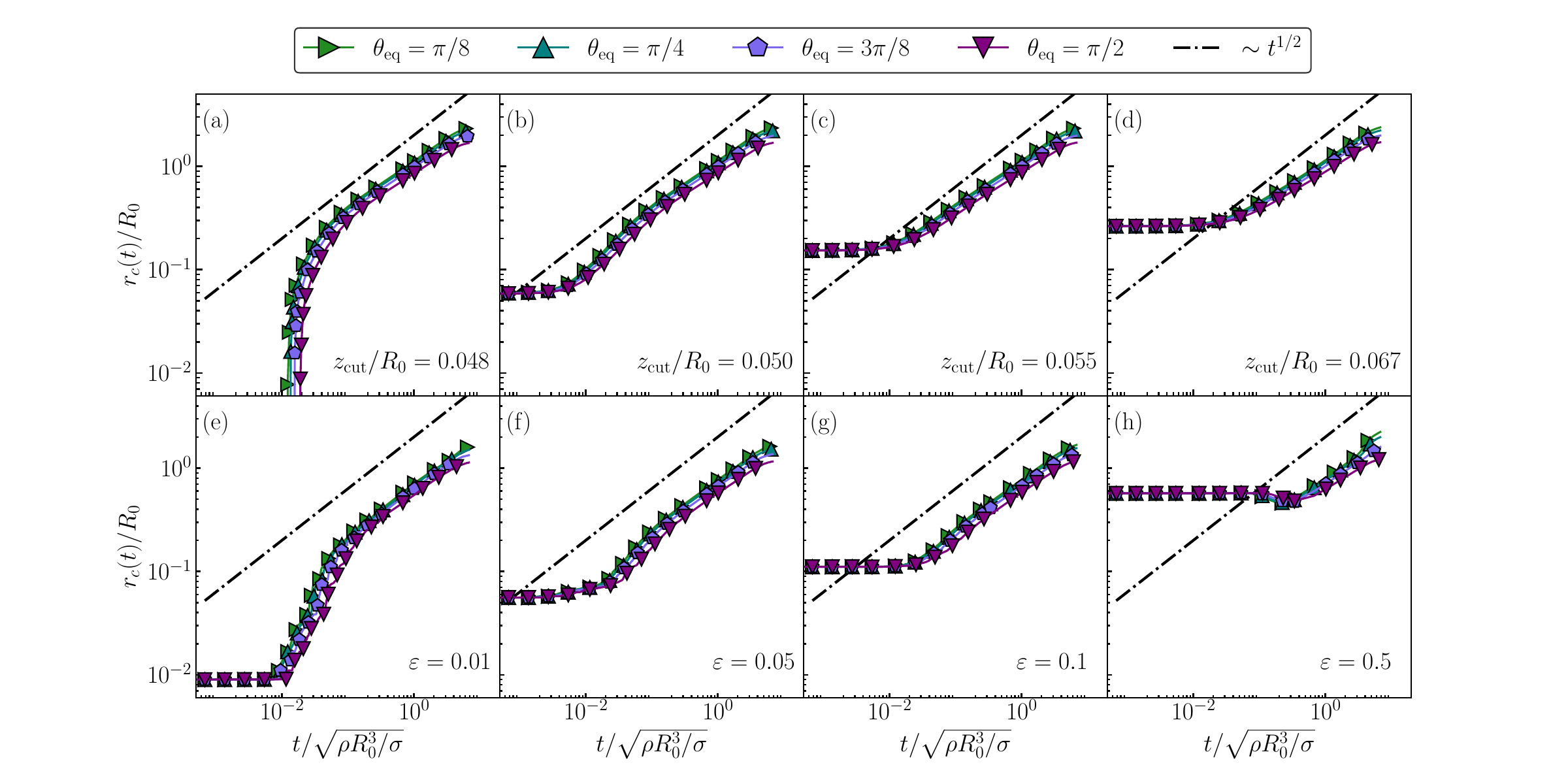}}%
    \caption{
    Droplet contact radius $r_c(t)$ as a function of time in the early stages of spreading dynamics, normalized with respect to the initial droplet radius $R_0$, for different equilibrium contact angles $\theta_{\text{eq}}$ (different colors/symbols). Reported time is normalized with respect to the characteristic inertial timescale $\tilde{t}$. To determine the proper definition of the contact radius, we compare the spreading dynamics for two definition protocols (see text for more details): top panels refer to protocol ({\it i}), based on the intersection parameter $z_{\rm cut}$, and show results for different values of $z_{\rm cut}/R_0$; bottom panels refer to protocol ({\it ii}), based on the flatness parameter $\varepsilon$, and show results for different values of $\varepsilon$.}
    \label{fig:inertial_scaling}
\end{figure*}
%%%%%%%%%%%%%%%%%%%%%%%%%%%%%%%%%%%%%%%%%%%%
In this section, we report on results of numerical simulations for the spreading dynamics of a droplet on a flat surface. The process of droplet spreading is characterized by different regimes, where different force contributions balance and produce different scaling laws for the evolution of the contact radius $r_c(t)$, i.e., the radius of the interface area that is in contact with the wall. The initial stages of spreading have been investigated in several experimental, numerical and theoretical works~\cite{biance2004first,ding2007inertial,Bird08,Courbin09,carlson2011dissipation,carlson2012universality,Winkels12, eddi2013short,chen2014effects,frank2015lattice, LegendreMaglio13,legendre2015comparison,jose2017role,baroudi2020effect,Pelusi22}. 
If Laplace pressure balances the inertial terms, an inertial scaling regime is expected, $r_c(t) \sim t^{1/2}$, echoing the inertial scaling of the contact area between two coalescing droplets~ \cite{aarts2005hydrodynamics,eggers1999coalescence,duchemin2003inviscid,wu2004scaling,thoroddsen2005coalescence,case2008coalescence,paulsen2011viscous}. This inertial scaling is verified in spreading experiments on completely wetting surfaces as well as in the initial stages of spreading of droplets on partially wetting surfaces, while for later stages a departure from the $1/2$ exponent is observed~\cite{biance2004first,Bird08,Courbin09,Winkels12}. 

In the very late stage of spreading dynamics, a much slower process is expected, when viscous effects balance surface tension forces, reproducing the celebrated Tanner's law, $r_c(t) \sim t^{1/10}$ ~\cite{Tanner79,voinov1976hydrodynamics,bonn2009wetting}. Since the wetting modeling implemented in our IB-LB method (see Sec.~\ref{subsec:wetting_interaction}) does not result in a direct contact between the interface and the wall, it is natural to look for an appropriate definition of the contact radius $r_c(t)$. For this reason, we first investigate the sensitivity of the expected scaling laws to the way $r_c(t)$ is defined. With this aim, we design two protocols to determine the contact radius: 
({\it i}) following an earlier work on the subject~\cite{chamakos2017design}, we identify the contact line as the intersection between the droplet interface and a plane parallel to the wall placed at the height $z_{\rm cut}$, referred to as the {\it intersection parameter}; then, we evaluate the contact radius as the average distance of the contact line from the center of the contact plane at height $z_{\rm cut}$; 
({\it ii}) we consider the portion of the droplet interface that forms a flat region parallel to the wall: since the droplet surface is discretized into triangles, one has to focus on the lower half of the droplet and consider each triangular element individually.  A triangle, whose vertices are positioned at $\vec{q}_i$ (see Sec.~\ref{subsec:IBLB}), is included as part of the flat region if $\max_{i,j} \vert (\vec{q}_i-\vec{q}_j)\cdot\vec{\hat{n}}_w \vert<\varepsilon R_0$, where the indices $i$ and $j$ run on the three vertices of the triangle, $\vec{\hat{n}}_w$ is the unit normal to the wall, and $\varepsilon$ is a threshold parameter, referred to as the {\it flatness parameter}. The contact area computed using this procedure ($\chi_{\varepsilon}$) is determined as the sum of the areas of all triangles that satisfy the flatness criterion. The corresponding contact radius, $r_{\varepsilon}$, is then calculated by assuming that $\chi_{\varepsilon}$ corresponds to the area of a circle with radius $r_{\varepsilon}$. 

%%%%%%%%%%%%%%%%%%%%%%%%%%%%%%%%%%%%%%%%%%%%
\begin{figure}[h]
    \centering
    \includegraphics[width=0.49\textwidth]{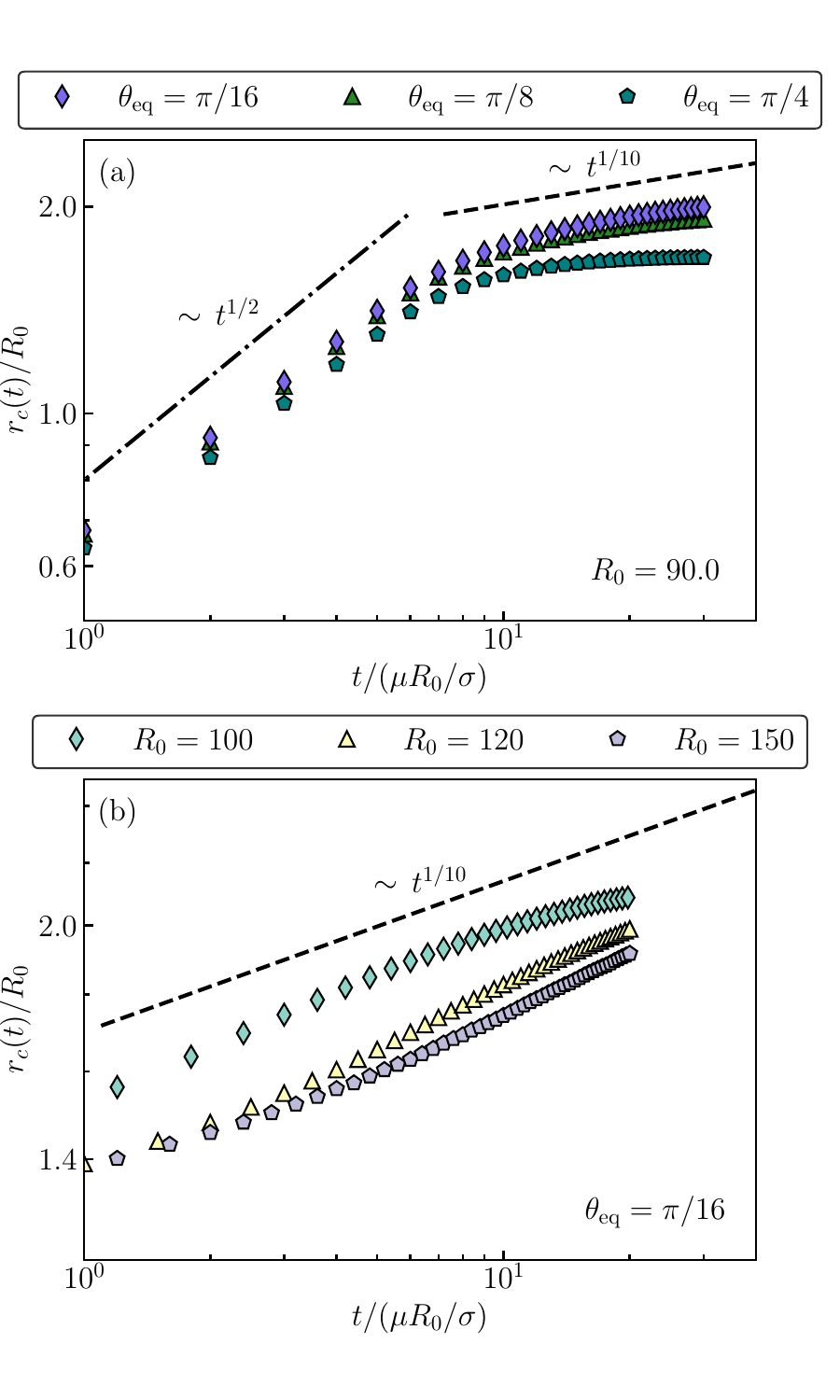}%
    \caption{Dynamics of the droplet contact radius $r_c(t)$ at early and late stages of spreading. 
    Contact radius is normalized with respect to initial droplet radius $R_0$. Time is normalized with respect to characteristic viscous time $t^*$.
    Top panel: evolution of $r_c(t)$ for different values of the equilibrium contact angle $\theta_{\text{eq}}$ (different colors/symbols) based on simulations with $R_0\!=\!90$; scaling laws for inertial regime (dash-dotted line) and viscous regime (dashed line) are also reported. 
    Bottom panel: evolution of the contact radius for a fixed contact angle ($\theta_{\rm eq}\!=\!\pi/16$) and different values of $R_0$ (different colors/symbols) based on simulations where the droplet is initialized as a spherical cap; dashed line refers to the scaling law for viscous regime.
    }
    \label{fig:drops_dynamics}
\end{figure}
%%%%%%%%%%%%%%%%%%%%%%%%%%%%%%%%%%%%%%%%%%%%

We first focus on the dynamics in the early stages of wetting, where an inertial scaling law $r_c(t)\!\sim\!t^{1/2}$ is expected for the contact radius $r_c(t)$. By using protocols ({\it i}) and ({\it ii}) and comparing against the expected scaling law, we assess the robustness in measuring $r_c(t)$ as the free parameters $z_{\rm cut}$ and $\varepsilon$ are varied. This was done by performing dedicated simulations with $\xi\!=\!2$, and $\lambda\!=\!10$ in a domain of size $\displaystyle L_x \!\times\! L_y \!\times\! L_z \!=\! 240 \!\times\! 240 \!\times\! 130$. Simulations are initialized as described in Sec.~\ref{subsec:STATICS}, with a spherical droplet of radius $R_0\!=\!60$ initially placed close to the wall, at distances where the wall-interaction force is not negligible in order to enable spreading. Notice that we chose a value of $R_0$ larger than the ones used for the static simulations to achieve a smaller Ohnesorge number, $\text{Oh}\!=\!\mu/(\rho R_0 \sigma)^{1/2}$, and facilitate comparisons with literature results. In Fig.~\ref{fig:inertial_scaling}, we analyze the time evolution of the contact radius $r_c(t)$ for both protocols ({\it i}) and ({\it ii}). The reported time is normalized with the inertial timescale $\tilde{t}\!=\!\sqrt{\rho R_0^3/\sigma}$~\cite{Bird08,Winkels12,LegendreMaglio13,du2021initial}. Figure \ref{fig:inertial_scaling}(a)-\ref{fig:inertial_scaling}(d) refer to protocol ({\it i}) and show different values of the normalised intersection parameter $z_{\rm cut}/R_0$, while Fig.~\ref{fig:inertial_scaling}(a)-\ref{fig:inertial_scaling}(d) report results obtained by employing protocol ({\it ii}) for different values of the flatness parameter $\varepsilon$. Different symbols/colors correspond to different values of the equilibrium contact angles $\theta_{\text{eq}}$. The inertial scaling $r_c(t)\!\sim\!t^{1/2}$ is also reported for comparison. 
As expected, varying the parameter that pertains to each protocol has a non-trivial impact on the measured radius, particularly at early stages when the contact radius is small.
For protocol ({\it i}), the plateau is seen until the points on the contact plane, fixed at a height $z_{\rm cut}$ from the wall, are unaffected by the spreading process.
For protocol ({\it ii}), a larger value of $\varepsilon$ means a triangle is considered flat with a lower tolerance; therefore, the region judged as flat is constant for a longer period. 
Then, when the contact radius starts to increase, it soon approaches an inertial-like scaling with a power-law behavior very close to $r_c(t)\!\sim\! t^{1/2}$: this happens for all the considered values of $z_{\rm cut}$ and $\varepsilon$, although its onset occurs at different times depending on the values of the two parameters, respectively. 
The earliest onsets for the two protocols is seen in Fig.~\ref{fig:inertial_scaling}(b) and Fig.~\ref{fig:inertial_scaling}(f), where it occurs around $t/\tilde{t}\!\sim \!0.01$, with the plateau for protocol ({\it i}) [Fig.~\ref{fig:inertial_scaling}(b)] lasting slightly less than the plateau for protocol ({\it ii}) [Fig.~\ref{fig:inertial_scaling}(f)]. 
Furthermore, it emerges that there exist optimal values of $z_{\rm cut}/R_0\!\approx\!0.050$ and $\varepsilon\!\approx\!0.1$ for which one can get closer to the inertial scaling, showing a nice agreement with molecular dynamics results reported in Ref.~\cite{Winkels12}. Interestingly, the optimal value of $z_{\rm cut}$ matches the height where we expect to have the maximum mean curvature for the droplet interface at equilibrium, which roughly occurs at the minimum of the wetting interaction term $\Pi(h\!=\!h^*)$, where $h^* \!=\! \xi (n/m)^{1/(n-m)}$. For the chosen values of $m$, $n$, and $\xi$, this leads to $h^*/R_0\!\approx\!0.050$. Regarding the precise details of the power-laws observed, although the corresponding exponents may trivially be fitted~\cite{du2021initial}, visual inspection reveals they are essentially very close to $1/2$. Some slight dependence on wettability is observed, especially for the larger values of $\varepsilon$, with data for smaller $\theta_{\text{eq}}$ lying slightly above data with larger $\theta_{\text{eq}}$, in agreement with the literature~\cite{Bird08,Winkels12,du2021initial}. Improving the quality of data below $r_c(t)/R_0\!=\!0.01$ would probably require larger resolutions. 
In summary, the measurement of the contact radius $r_c(t)$ sensibly depends on the specific values of the parameter involved in the protocol used. Moreover, there exists an optimal value of the parameter for which the agreement with the expected scaling law is maximized.
Although in Fig.~\ref{fig:inertial_scaling} it emerges that the choice of the protocol does not markedly impact the optimal agreement with theory, protocol ({\it i}) could be favored, since it provides a physically intuitive relation that connects the optimal value of the intersection parameter $z_{\rm cut}$ and the height of the maximum mean curvature.

Next, we investigate the late stage of the dynamics, which is expected to be dominated by viscous forces. 
In Fig.~\ref{fig:drops_dynamics}(a), we report results obtained by performing longer simulations with $\xi\!=\!3$, $R_0\!=\!90$, $N_t\!=\!500\,000$ and different equilibrium contact angles $\theta_{\text{eq}}\!\in\! \{\pi/16,\, \pi/8,\, \pi/4\}$. 
For larger values of $\theta_{\text{eq}}$, the droplet hardly spreads before reaching equilibrium after a brief inertial-spreading stage.
For small values of $\theta_{\text{eq}}$, however, before the spreading stops, we observe a transition to another scaling close to $r_c(t)\!\sim\!t^{1/10}$, as expected from Tanner's law~\cite{Tanner79,voinov1976hydrodynamics,bonn2009wetting}, which arises when viscous and surface tension forces prevail.
This observation well aligns with previous works~\cite{LegendreMaglio13,du2021initial}.
To study the scaling law behavior of the late stage in more detail, we performed additional simulations using a dedicated set-up: the droplet is initialized as a spherical cap, bypassing the initial inertia-dominated spreading process, allowing us to focus directly on the viscous scaling law. Simulations are performed by fixing $\xi$ and using different $R_0\!\in\!\{100,\,120,\,150\}$ using $N_t\!=\!500\,000$ for the former two radii and $N_t\!=\!\,750\,000$ for the last one.  
The results are reported in Fig.~\ref{fig:drops_dynamics}(b). We observe that, by increasing $R_0$, the scaling law $r_c(t) \sim t^{1/10}$ is more closely followed, suggesting that a clear separation between the (inner) regularizing lengthscale $\xi$ and the (outer) lengthscale $R_0$ is necessary to observe Tanner's law. 
Interestingly, the prefactor is a decreasing function of $R_0$: when studying viscous spreading modeled using sharp-interface hydrodynamics with a slip length $\ell_s$, the predicted prefactor scales like $\sim\!1/[\log (R_0/\ell_s)]^{1/10}$. This suggests that, when $R_0/\xi \!\gg \!1$, the mechanism of regularization introduced by the finite lengthscale $\xi$ 
effectively plays the role of a slip boundary condition on the substrate.
More detailed comparisons with the spreading process of a droplet are needed to quantitatively address this point.

%%%%%%%%%%%%%%%%%%%%%%%%%%%%%%%%%%%%%%%%%%%%
\section{Conclusions}\label{sec:CONCLUSIONS}
%%%%%%%%%%%%%%%%%%%%%%%%%%%%%%%%%%%%%%%%%%%%
We have studied the statics and dynamics of wetting within the framework of the hybrid immersed boundary - lattice Boltzmann (IB-LB) numerical technique, focusing on the problem of the interaction of a liquid droplet with a solid wall. We leveraged a previous IB-LB model for wetting by some of the authors~\cite{Pelusi2023}, with the main scope to overcome its limitations related to the difficulty in modeling small contact angles and the need to use pre-calibration simulations to set the desired contact angles. We have substantially extended and improved the method in various directions. We used a controlled implementation of wetting~\cite{Chamakos_2013,Karapetsas_2016,du2021initial} featuring a wall-interaction term $\Pi(h)$ with strong repulsions at short distances and attraction at intermediate distances, thus impacting the profile shape only in the vicinity of the wall. Within this approach, the equilibrium contact angle can be set {\it a priori} via the suitable choice of the $\Pi(h)$ parameters. Furthermore, a regularization lengthscale $\xi$ is present, which allows to control abrupt curvature changes close to the contact line. Importantly, at sufficiently large distances from the wall, the macroscopic view of wetting is recovered, with the static interface shape approaching a wedge-like structure with the desired equilibrium contact angle $\theta_{\text{eq}}$. 
Theoretical predictions for the equilibrium shapes have been successfully verified in simulations.
Moreover, we have further used simulations to investigate the dynamics of spreading, both in the inertia-dominated and in the viscous-dominated regimes. The results have been found to be in agreement with those in the known literature.

Our results pave the way for interesting future developments. First of all, the IB-LB methodology has been widely and successfully used to model complex interface physics, e.g. featuring elasticity and interface viscosity~\cite{Guglietta2020,rezghi2022tank,liSimilarDistinctRoles2021,guglietta2020lattice,guglietta2021loading,guglietta2023suspensions,rezghi2022lateral,li2020finite,guglietta24analytical,liFiniteDifferenceMethod2019}. Studying wetting problems involving interface physics with these additional complexities surely poses excellent food for thought in the future perspective. For example, many studies focused on elastic wetting, featuring liquids close to solid walls confined by elastic membranes~\cite{hosoi2004peeling,lister2013viscous,carlson2018fluctuation,poulain2022elastohydrodynamics,saeter2024coalescence}. The IB-LB is particularly amenable in accommodating elasticity models at the interface via a suitable change in the constitutive law: this amounts to considering different strain energies in Eq.~\eqref{eq:force_bb}, embedding more information than just surface tension forces; hence, it could be used as an ideal numerical tool to complement experimental observations and/or extend our knowledge on wetting to unexplored settings. Another key noteworthy aspect is the fact that the IB-LB method is particularly efficient; hence, it can be used to generate ground-truth data and explore the possibility of developing data-driven methodologies to unravel the non-linear relation between the droplet dynamics and the features that control it ~\cite{demou2024hybrid,demou2023ai}.

\begin{acknowledgments}
The authors acknowledge Luca Biferale for insightful preliminary discussions.
This research is supported by European Union’s HORIZON MSCA Doctoral Networks programme, under Grant Agreement No. 101072344, project AQTIVATE (Advanced computing, QuanTum algorIthms and data-driVen Approaches for science, Technology and Engineering). 
FP and MS acknowledge the support of the National Center for HPC, Big Data and Quantum Computing, Project CN\_00000013 – CUP E83C22003230001 and CUP B93C22000620006, Mission 4 Component 2 Investment 1.4, funded by the European Union - NextGenerationEU. 
FG acknowledges the support of the Italian Ministry of University and Research (MUR), FARE program (No. R2045J8XAW), project ``Smart-HEART''.
Funding  from the European Union’s Horizon 2020 research and innovation programme under grant agreement Nos 882340 (European Research Council) and 810660 is also acknowledged. 
\end{acknowledgments}

\appendix

%%%%%%%%%%%%%%%%%%%%%%%%%%%%%%%%%%%%%%%
\section{Setting the wetting interaction strength \texorpdfstring{$A$}{Lg}}\label{appendix:setting_angle}
\begin{figure}
    \centering
    \includegraphics[scale=1]{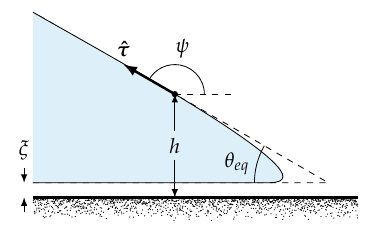}
    \caption{Wedge-like profile in the limit $\xi/R_0\to0$, which is used for obtaining the dependence of $A$ as a function of $\theta_{\text{eq}}$.\vspace*{-4mm}}
    \label{fig:wedge_geometry}
\end{figure}

In this section, we provide details on how to relate the constant $A$ in the wetting interaction term in Eq.~\eqref{eq:disjoin-p} to the equilibrium contact angle $\theta_{\text eq}$ in the limit of large droplets.  At equilibrium (${\bm U}={\bm 0}$), the force balance equation in the normal direction reduces to Eq.~\eqref{eq:laplace-pi}. For large droplets, the scale $\xi$ is much smaller than  the characteristic droplet radius $R_0$ and we also have $\Delta p\to 0$ as a consequence of the fact that the pressure must be inversely proportional to $R_0$. Therefore, in the limit when $R_0\to\infty$, Eq.~\eqref{eq:laplace-pi} reduces to $\sigma {\bm \nabla} \cdot \bm{\hat{n}}+\Pi(h)\approx 0$ and the free-surface assumes a wedge-like  profile in the vicinity of the contact line, whose size allows us to neglect its curvature in the azimuthal direction (see Fig.~\ref{fig:wedge_geometry}). In this manner, the curvature of the profile may be obtained via the Serret--Frenet formulas \cite{Pressley2010} 
%%%%%%%%%%%%%%%%%%%%%%%%%%%%%%%%%%%%%%%

\begin{figure}
    \centering
    \includegraphics[scale=1]{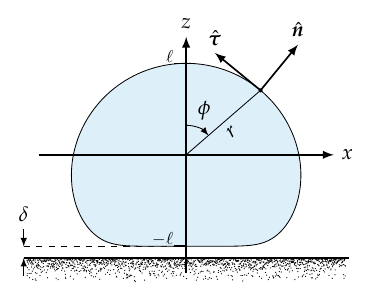}
    \caption{Droplet geometry. The droplet is centred around the vertical axis and lies at a distance $\delta$ above the substrate.}
    \label{fig:drop_geometry}
\end{figure}
%%%%%%%%%%%%%%%%%%%%%%%%%%%%%%%%%%%%%%%%%%%%%%%%%%
\begin{equation}
    {\bm \nabla} \cdot \bm{\hat{n}}=\left\| \frac{\mathrm{d}\bm{\hat{\tau}}}{\mathrm{d}s}\right\|=\frac{\mathrm{d}\psi}{\mathrm{d}s},
\end{equation}
where $s$ is arc length, $\bm{\hat{\tau}}$ is the tangent vector and $\psi$ is the angle it makes with the $x$-axis  (see  Fig.~\ref{fig:wedge_geometry}). Since the $z$-coordinates of the wedge-like profile, $h(s)$, are linked with $\psi(s)$ through $\sin\psi = \mathrm{d}h/\mathrm{d}s$ \cite{Boucher1980}, we may invoke the chain rule to write Eq.~\eqref{eq:laplace-pi} in the large-droplet limit as
\begin{equation}\label{eq:psiPi}
   \sigma \sin\psi \frac{\mathrm{d}\psi}{\mathrm{d}h} + \Pi(h) = 0.
\end{equation}
To show Eq.~\eqref{eq:A_choice}, we use the fact that Eq.~\eqref{eq:psiPi} is a separable differential equation and we perform integrals over the range $h$ and $\psi$ are defined, namely from $\psi\!=\!0$ (equivalently $h\!=\!\xi$ so that Eq.~\eqref{eq:psiPi} holds) to $\psi\!=\!\pi-\theta_{\text{eq}}$ [equivalently $h\to\infty$, see Fig.~\ref{fig:wedge_geometry}], so that
\begin{equation}
\sigma \int_0^{\pi-\theta_{\text{eq}}}\sin\psi\, \mathrm{d} \psi =-A \int_{\xi}^\infty  \left[\left( \frac{\xi}{h}\right)^n-\left( \frac{\xi}{h}\right)^m\right]\mathrm{d} h.
\end{equation}
The above integrals may be straightforwardly evaluated to yield Eq.~\eqref{eq:A_choice}~\cite{Chamakos_2013,Karapetsas_2016,du2021initial}. 

%%%%%%%%%%%%%%%%%%%%%%%%%%%%%%%%%%%%%
%%%%%%%%%%%%%%%%%%%%%%%%%%%%%%%%%%%%%
\section{Solving for droplet shapes}\label{appendix:droplet_shape}
%%%%%%%%%%%%%%%%%%%%%%%%%%%%%%%%%%%%%

In this section, the main ingredients which are necessary to obtain numerical solutions to Eq.~\eqref{eq:laplace-pi} are provided. A key step towards this goal is casting the surface profile as a singe-valued function, which is conveniently accomplished by considering the droplet in a spherical coordinate system. Assuming axisymmetry, the radial function identifying the surface is  $r=G(\phi)$, where $\phi$ is the polar angle formed with the $z$-axis (see Fig.~\ref{fig:drop_geometry}). This notation for the polar angle has been decided against the popular choice of $\theta$ to avoid notation clash with the contact angle which uses the same variable. The problem is made non-dimensional using $G(\phi)\!=\!R_0g(\phi)$, where, like previously, $R_0$ is the initial (spherical) droplet radius. This particular form for the droplet profile avoids numerical issues related to the discretization of surfaces of different sizes. The outward  unit normal of the surface, $\hat{\boldsymbol{n}}$ is given by
\begin{equation}
 \hat{\boldsymbol{n}} =\frac{g}{(g^2+g^2_{\phi})^{1/2}} \hat{\boldsymbol{r}} - \frac{g_{\phi}}{(g^2+g^2_{\phi})^{1/2}} \hat{\boldsymbol{\phi}},
 \end{equation}
where $\hat{\boldsymbol{r}} $ and $\hat{\boldsymbol{\phi}}$ are the unit vectors in the radial and polar directions, respectively, with subscripts denoting partial differentiation with respect to the denoted variable, $\phi$ here. Evaluating the divergence of the unit normal in spherical coordinates, we find 
\begin{equation}
    {\bm \nabla} \cdot \bm{\hat{n}} = \frac{2 g^2+3 g^2_{\phi}-g g_{\phi \phi}}{R_0(g^2+g^2_{\phi})^{3/2}}-\frac{g_{\phi} \cos \phi}{R_0g \sin \phi (g^2+g^2_{\phi})^{1/2}}.\label{eq:divn}
\end{equation}
To facilitate the numerical solution in the vicinity of the polar axes, we notice that $g(\phi)$ is in fact a function of $\cos\phi$. Making the variable change $u\!=\!\cos\phi$, Eq.~\eqref{eq:divn} is cast as an equation for $g(u)$, namely
\begin{equation}
    {\bm \nabla} \cdot \bm{\hat{n}}  = \frac{ ug\dot{g} - g^2 - (1-u^2)g\ddot{g}}{R_0\left[(1-u^2)\dot{g}^2+g^2\right]^{3/2}} +     \frac{u \dot{g} + 3g }{R_0g \sqrt{(1-u^2)\dot{g}^2+g^2}},
\end{equation}
where the dots denote differentiation with respect to $u\!\in\![-1,1]$. The normal force balance equation~\eqref{eq:laplace-pi} becomes: 
\begin{align}\label{eq:YL}
 k&=\frac{ ug\dot{g} - g^2 - (1-u^2)g\ddot{g}}{\left[(1-u^2)\dot{g}^2+g^2\right]^{3/2}}  
    + \frac{u \dot{g} + 3g }{g \sqrt{(1-u^2)\dot{g}^2+g^2}}  \nonumber\\
    &+ \frac{(m-1)(n-1)}{(n-m)\epsilon}(1+\cos\theta_{\text{eq}})\left[\left( \frac{\epsilon}{h}\right)^n\!\!-\left( \frac{\epsilon}{h}\right)^m\right],
\end{align}
where we set $\epsilon\!=\!\xi/R_0$, $k\!=\!\Delta p R_0/\sigma$ and $h$ is the distance from the substrate given by $h\!=\! u g(u)+g(-1)+\delta$, with $\delta\approx \epsilon$ being the smallest distance from the substrate and needs to be determined as part of the solution (see Fig.~\ref{fig:drop_geometry}). Hence, given $k$ (or equivalently, the volume of the droplet, see below), we determine $\delta$ and $g(u)$, subject to
\begin{equation}\label{eq:bc}
g(1)=g(-1),
\end{equation}
which means that the origin is placed in the middle between the top and bottom parts of the droplet. Through an integral constraint, we impose the volume of the droplet, which is written as
\begin{equation}\label{eq:volconstr2}
\int_0^\pi g^3(\phi)\sin\phi\,\mathrm{d}\phi=2,
\end{equation}
or, in terms of the $u$ variable, as
\begin{equation}\label{eq:volconstr}
\int_{-1}^1 g^3(u)\,\mathrm{d}u=2.
\end{equation}
Hence, to determine the droplet profile together with $\delta$ and $k$, we use second-order finite differences to enforce Eq.~\eqref{eq:YL} everywhere in the domain alongside with conditions Eqs.~\eqref{eq:bc} and \eqref{eq:volconstr}, thus maintaining the same number of degrees of freedom in this process. 

%%%%%%%%%%%%%%%%%%%%%%%%%%%%%%%%%%%%%%%%%%%
\section{Asymptotic structure of equilibrium solutions}\label{appendix:DROPLET_ASYMPTOTICS}
%%%%%%%%%%%%%%%%%%%%%%%%%%%%%%%%%%%%%%%

The present section scrutinizes equilibrium solutions determined from Eq.~\eqref{eq:YL} analytically in order to uncover further insight into their underlying structure. It is clear that equilibrium solutions consist of a nearly flat region near the wall for which $u\to -1$ (equivalently $\phi\to\pi$) and a spherical cap away from the wall and as $u\to1$ (equivalently $\phi\to0$). This observation allows us to approximate the solution within their respective regions.

Far from the wall, the interaction term $\Pi(h)$ in Eq.~\eqref{eq:YL} is generally small, particularly when $\epsilon=\xi/R_0$ is also small. When $\Pi(h)$ is absent, the solution to Eq.~\eqref{eq:YL} is exactly described by a cap of a sphere of radius $R_c = cR_0$, for some constant $c$. Since in these cases we know that the pressure jump satisfies $\Delta p = 2\sigma/R_c$, we readily conclude from the definition of $k$ in Appendix~\ref{appendix:droplet_shape} that $c = 2/k$. From simple geometrical arguments concerning the droplet size, the way the coordinate system is defined (see Fig.~\ref{fig:drop_geometry}) and the requirement that the droplet meets the substrate at the equilibrium angle $\theta_{\text{eq}}$, we readily obtain that the droplet profile is essentially that of a circle of radius $R_c$ which is shifted by a distance $R_c\cos^2(\theta_{\text{eq}}/2)$ downwards along the $z$-axis with a contact line of radius:
\begin{equation}\label{eq:contactr}
    r_c=R_c\sin \theta_{\text{eq}}. 
\end{equation}
Starting from the Cartesian representation of the circle 
\begin{equation}
    x^2 + y^2 + \left(z+R_c\cos^2\frac{\theta_{\text{eq}}}{2}\right)^2 = R_c^2
\end{equation}
and converting it to the spherical coordinate system, we find that the droplet surface is described by
\begin{equation}
    r = R_c\left[-\cos^2\frac{\theta_{\text{eq}}}{2}\cos\phi + \sqrt{1-\cos^4\frac{\theta_{\text{eq}}}{2}\sin^2\phi}\right], 
\end{equation}
noting that, as expected, the corresponding $g(u)$, namely
\begin{equation}\label{eq:gcap}
    g(u) = c\left[-\cos^2\frac{\theta_{\text{eq}}}{2}u + \sqrt{1-\cos^4\frac{\theta_{\text{eq}}}{2}(1-u^2)}\right], 
\end{equation}
satisfies Eq.~\eqref{eq:YL} with $k=2/c$ when the $\Pi(h)$ term is not present. Based on these considerations, the half-height of the droplet, $\ell$, is estimated as (see Fig.~\ref{fig:drop_geometry}) 
\begin{equation}\label{eq:ell}
    \ell = R_c\sin^2\frac{\theta_{\text{eq}}}{2},
\end{equation}
where corrections of order $\epsilon$ have been neglected. 

Given that this model dictates smooth variations in curvature near the contact line with larger $\epsilon$ accommodating more gentle curvature variations, defining the contact angle cannot be done in a consistent manner. However, by invoking the spherical cap approximation, one can estimate the corresponding $\theta_{\text{eq}}$ had $\epsilon$ been vanishingly small. This can be accomplished by utilizing the half-height of the droplet, which can be more accurately measured. Specifically, starting from the formula for the volume of a spherical cap in terms of $\tan (\theta_{\text{eq}}/{2})$ (see Appendix in~\cite{demou2024hybrid}), we deduce that
\begin{equation}\label{eq:voldrop}
    \frac{8R_0^3}{r_c^3}= \tan \frac{\theta_{\text{eq}}}{2} \left(3+\tan^2 \frac{\theta_{\text{eq}}}{2}\right).
\end{equation}
Then, by combining Eq.~\eqref{eq:contactr} with Eq.~\eqref{eq:voldrop}, we obtain an expression for $R_c$ in terms of $R_0$ and $\theta_{\text{eq}}$, which, when substituted in Eq.~\eqref{eq:ell} yields
\begin{equation}
\tan^2\frac{\theta_\text{eq}}{2} =  \frac{3\ell^3}{R_0^3-\ell^3}.\label{eq:angleh}
\end{equation}
\begin{figure}
\centering\includegraphics[scale=1]{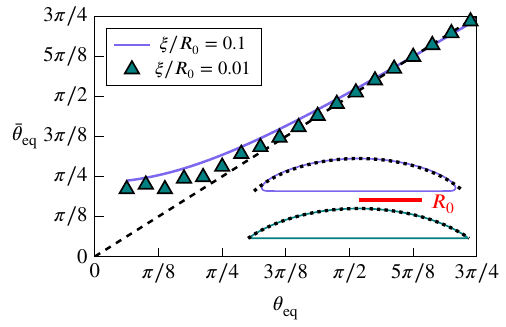}
    \caption{Measured equilibrium contact angle, $\bar{\theta}_{\text{eq}}$, using the height of the droplet and Eq.~\eqref{eq:angleh}, as a function of the imposed one, $\theta_{\text{eq}}$, for $\xi/R_0=0.1$ (solid curve) and $\xi/R_0=0.01$ (symbols) compared with $\theta_{\text{eq}}=\bar{\theta}_{\text{eq}}$ (dashed line), which is expected to hold as $\xi/R_0\to0$. The inset shows droplet profiles for $\theta_{\text{eq}}=\pi/4$ in the cases of $\xi/R_0=0.1$ (top; $\bar{\theta}_{\text{eq}}\approx0.32\pi$) and $\xi/R_0=0.01$ (bottom; $\bar{\theta}_{\text{eq}}\approx0.28\pi$). The red-colored line is a scale bar for $R_0$ which is used for scaling the profiles shown; the dotted curves  correspond to fitted circular arcs matching the height and volume of the drop.}
    \label{fig:angles}
\end{figure}%
Thus, the angle measured using Eq.~\eqref{eq:angleh} may be used to compare with the value of $\theta_{\text{eq}}$ imposed in Eq.~\eqref{eq:YL} as a means to assess the validity of the simplifying arguments invoked in Appendix~\ref{appendix:setting_angle}. Figure~\ref{fig:angles} shows a plot $\theta_{\text{eq}}$ measured using Eq.~\eqref{eq:angleh} as a function of the imposed angle $\pi/16\le \theta_{\text{eq}}\le 3\pi/4$, and for  $\epsilon=\xi/R_0=0.1$ and $0.01$. As expected, agreement improves as $\epsilon\to0$, but it remains generally better for obtuse angles compared to acute ones. The reason is that for smaller angles, the droplet's height decreases, making the assumption in Appendix~\ref{appendix:setting_angle} of an infinitely large wedge less relevant in this limit. Although the analysis in the Appendix could be revisited to account for finite-size effects, we chose not to pursue this approach, as the correction terms found did not significantly affect the results in the axisymmetric setting considered here.

\begin{figure}[!hb]
\centering\includegraphics[scale=1]{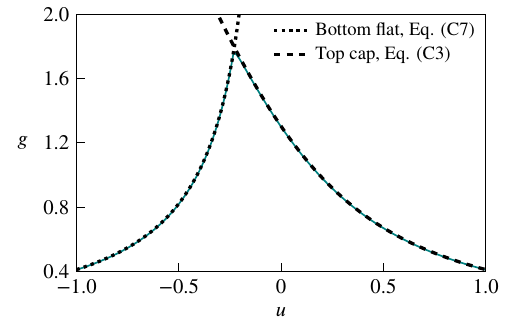}
    \caption{Plot of $g$ as a function of $u$ when $\xi/R_0=0.01$ and $\theta_{\text{eq}}=\pi/4$. Solid green line refers to the solution of Eq.~\eqref{eq:YL}. Dotted and dashed curves correspond to the flattened and cap-shaped regions of the droplet given by Eqs.~\eqref{eq:gcap} and \eqref{eq:gplanar}, respectively, using the measured angle $\theta_{\text{eq}}\approx0.28\pi$.}
    \label{fig:asymptotics}
\end{figure}%

Also noteworthy is that Eq.~\eqref{eq:YL} admits planar solutions. The flattened part of the droplet, is assumed to lie on the plane $z=-\ell$ (see Fig.~\ref{fig:drop_geometry}), which is expressed in spherical coordinates as  $r=-R_c\sin^2(\theta_{\text{eq}}/2)/\cos\phi$ [see Eq.~\eqref{eq:ell}]. Equivalently, using $r=R_0g(u)$ with $u=\cos\phi$, we find that $g$ is approximated well as $u\to-1$ by
\begin{equation}\label{eq:gplanar}
    g(u) = -\frac{c \sin^2 \dfrac{\theta_{\text{eq}}}{2}}{u}.
\end{equation}
Since planar solutions have no curvature ($\bm\nabla\cdot\hat{\boldsymbol{n}}=0$),  Eq.~\eqref{eq:YL} becomes
\begin{equation}
    \left( \frac{\epsilon}{\delta}\right)^n-\left( \frac{\epsilon}{\delta}\right)^m =\frac{(n-m)\epsilon k}{(m-1)(n-1)(1+\cos\theta_{\text{eq}})}=\beta,
\end{equation}
noting that $\delta\neq\epsilon$ is the distance from the substrate, see Fig.~\ref{fig:drop_geometry}). Solving this equation analytically is generally not possible. However we expect that $\delta<\epsilon$ and, in the limit of large drops ($\beta\to0$), we have that $\delta\to\epsilon^-$. However, for certain combinations of $n$ and $m$, an analytical solution may be found. For the combination of exponents used in the present study, $(n,m)=(6,3)$, we find
\begin{equation}
    \delta = \epsilon\left(\frac{2}{1+\sqrt{1+4\beta}}\right)^{1/3}.
\end{equation}
Figure~\ref{fig:asymptotics} compares the solution to Eq.~\eqref{eq:YL} with the approximations given in Eqs.~\eqref{eq:gcap} and \eqref{eq:gplanar} for $\theta_{\text{eq}}= \pi/4$ and $\xi/R_0=0.01$. An excellent agreement is observed when the angle used is not the one imposed in the model, but the one measured using Eq.~\eqref{eq:angleh}, found to be $\theta_{\text{eq}}\approx0.28\pi$.

\bibliographystyle{apsrev4-2}
\bibliography{biblio}% Produces the bibliography via BibTeX.

\end{document}